\documentclass[pdflatex,sn-mathphys-num]{sn-jnl}% Math and Physical Sciences Numbered Reference Style
\usepackage{graphicx}
\usepackage{dcolumn}
\usepackage{bm}
\usepackage{physics} 
\usepackage[utf8]{inputenc}
\usepackage[T1]{fontenc}
\usepackage{mathptmx}
\usepackage{etoolbox}
\usepackage{xcolor}
\usepackage{amsmath}
\usepackage{amssymb}
\usepackage{caption}
\usepackage{subcaption}
\usepackage{float}
\usepackage{tikz}
\usepackage{hyperref}
\usepackage{graphicx,booktabs,array}
\usepackage{stackengine}
\usepackage{mathtools}

\newcommand{\dbr}{\text{d}\mathbf{r}}

\newcommand{\br}{\mathbf{r}}
\newcommand{\dm}{\rho_{1}}

\newcommand{\gfrrp}{G(\mathbf{r}, \mathbf{r}'; \beta)}

\newcommand{\gfrpr}{G(\mathbf{r}', \mathbf{r}; \beta)}

\newcommand{\gfxxp}{G(x,x'; \beta)}

\newcommand{\ham}{\hat{\mathcal{H}}}

\newcommand{\ef}{\epsilon_F}

\theoremstyle{thmstyleone}%
\theoremstyle{thmstyletwo}%

\theoremstyle{thmstylethree}%

\begin{document}

\title{ An investigation into the nonconformity of homogeneous gas limit for kinetic energy density of atomic systems}


\author[]{\fnm{Priya} \sur{Priya}}\email{priyam20@iitk.ac.in}
\author[]{\fnm{Anuvab} \sur{Panda}}\email{anuvabp25@iitk.ac.in}
\author[]{\fnm{Saswata} \sur{Basu}}\email{saswatabasu25@iitk.ac.in}
\author*[]{\fnm{Mainak} \sur{Sadhukhan}}\email{mainaks@iitk.ac.in}
\affil{Department of Chemistry, Indian Institute of Technology Kanpur, Uttar Pradesh, India}



% \affil*[1]{\orgdiv{Department}, \orgname{Organization}, \orgaddress{\street{Street}, \city{City}, \postcode{100190}, \state{State}, \country{Country}}}

% \affil[2]{\orgdiv{Department}, \orgname{Organization}, \orgaddress{\street{Street}, \city{City}, \postcode{10587}, \state{State}, \country{Country}}}

% \affil[3]{\orgdiv{Department}, \orgname{Organization}, \orgaddress{\street{Street}, \city{City}, \postcode{610101}, \state{State}, \country{Country}}}

\abstract{Developing a reliable kinetic energy density functional within orbital-free density functional theory remains a long-standing challenge, particularly for atomic and molecular systems. A major difficulty lies in the absence of a systematic approach to accurately compute the kinetic energy density in such contexts. In our recent work, we introduced an analytical Green's function-based framework to address this issue. The majority of the existing efforts to construct an approximate kinetic energy density for atomic systems use homogeneous electron gas as the bedrock of their formalism. In this work, we have shown by using a P\"oschl-Teller potential that for realistic atomic potentials such a model yields improper results, emphasizing the need to change the leading-order term for the quest for kinetic energy densities of atoms and molecules.   }




\maketitle

\section{Introduction}\label{intro}
% Write KED is bottleneck in OFDFT for atoms and molecules 
Kinetic energy density (KED) remains a central challenge  in the development of 
orbital-free density functional theory (OF-DFT), especially for atoms and molecules. One of the key difficulties lies in the absence of a general and systematic framework for the accurate approximation of KED in such systems. In this present work, using P\"oschl-Teller potential as a model for atomic systems, we have shown that the approximate methods that begin with pristine Thomas-Fermi kinetic energy as the leading contribution appear to be inadequate for atomic systems using an amalgamation of Hamiltonian partitioning scheme and a novel Green's function technique. Our choice of such a one-dimensional potential is motivated by the very recent developments \cite{theophilou_2018, nagy_potfunctional_21, nagy_dft_2024} proving that the spherically averaged electron density around the atomic nuclei uniquely determines the system potential. Since such potentials, barring very close to the nuclear positions, do have similar qualitative features of P\"oschl-Teller potential, we expect that our results will hold up for realistic atoms and molecules.

%Recently, we have developed an analytical formalism using the Green's function (GF), which is both general and systematic. It is sa pin-multiplicity-independent perturbative formalism to compute the KED for atomic and molecular electron densities in their ground electronic state. In this work, we have shown using a P\"oschl-Teller potential that for real atomic systems, the widely used homogeneous electron gas (HEG) as the leading order yields improper results. This finding offers a key insight into the inadequacy of HEG-based terms and emphasizes the need for alternative foundations when constructing KED functionals for atoms and molecules.\\
%has become widely employed in materials science and solid-state chemistry, largely due to the development of effective kinetic energy density (KED) functionals. In solid-state systems, where electron densities are relatively homogeneous and approach the Thomas-Fermi limit, these functionals perform reliably \cite{carter_2012_chem_rxn, carter_PRB2010, carter_2018}. However, in atomic and molecular systems, the pronounced inhomogeneity of the electron density presents significant challenges. Developing an accurate approximation of KED functionals that can accurately capture both global and local electronic properties in such systems remains an open and active area of research within the orbital-free DFT framework.\\
% work of Burke 
Diverse approximate frameworks have been developed over the years to construct an accurate approximation of  KED functionals. The earliest approximation of KE was modeled after the homogeneous electron gas (HEG) \cite{thomas1927calculation, fermi1927statistical} 
\begin{equation}
    T_{TF}[\rho] = C_{TF}\int  \rho^{5/3}(\br)d\br; \, C_{TF} = \left(\frac{3 \hslash^2}{10m}\right)\left(3 \pi^2\right)^{2/3}.
\end{equation}
To account for mild inhomogeneities in the electron density,   von Weizs\"acker  introduced  a gradient correction 
\cite{weizsacker1935theory}
\begin{equation}
    T_{vW} [\rho] = \frac{\hslash^2}{8m}\int  \frac{\abs{\nabla \rho(\br)}^2}{\rho(\br)} d\br.
\end{equation}
 The $T_{vW}$ is the exact kinetic energy (KE) for electrons in $1s$ orbital of a hydrogen atom. Despite their historical importance quite early on, these functionals, separately as well as in conjunction, have been proven inadequate for describing molecular binding \cite{Teller_RevModPhys.34.627}.  
To improve upon this theory further, Thomas–Fermi–Dirac–von Weizsäcker (TFDvW) theory was proposed, which combines the Thomas-Fermi term with scaled von Weizs\"acker correction \cite{ParrYang1994, yang_parr_1986}. While such approaches can be effective in extended systems like solids, they often fail to capture key local quantum features—such as shell structure in atoms and molecules. More recently, semiclassical expansion in the large-system limit using model systems like one-dimensional P\"oschl-Teller slabs to systematically improve upon the Thomas–Fermi theory\cite{Burke_2024_ptslab}. These approaches have shown promise for orbital-free simulations of materials, especially in planar or slab-like geometries. Unfortunately, their applicability to atomic systems with localized electron density remains limited. 
Interestingly, these studies have demonstrated that even simple one-dimensional models can offer valuable insights into complex orbital-free frameworks \cite{burke_chap7}. For instance, a proof-of-concept machine-learning-based framework, developed using 1D systems, has shown that such models can be extended toward constructing accurate KE functionals for real atomic environments \cite{Burke_2013_ML_1D}.\\
% It's intriguing how simple 1 dimesional systems can provide great insights about the complex framework and the results can be extended to real systems.
%\red{In this direction, it has also been demonstrated through a proof-of-concept study on simple one-dimensional that machine learning-based framework may be extended to real atomic systems, offering an alternative route for constructing accurate KE functionals \cite{Burke_2013_ML_1D}. } \\
Orbital-free density functional theory enjoys better success for solid-state systems due to the development of a class of KED functionals. These functionals such as the Wang–Govind–Carter (WGC) functional, combine local and gradient-based contributions with a non-local term. It is defined as $T_{WGC}[\rho] = T_{TF}[\rho]+ T_{vW}[\rho]+T_{NL}[\rho]$ where $T_{NL}$ is called the non-local KE
 \cite{carter_2018, carter_wang1999orbital, Carter_Ligneres2005, Carter_HUNG2009163,carter_2012_chem_rxn,carter_PRB2010,carter_2018}. 
 The non-local functionals and related functionals, often based on generalized gradient approximations or kernel-based corrections, have formed the foundation of many successful OF-DFT implementations for materials modeling \cite{luo2018, wesolowski1997kohn, WesolowskiWang2013, wesolowski1993frozen, wesolowski1997density}.In contrast, the electron density in atoms and molecules is far more inhomogeneous compared to that in bulk materials. Consequently, the enhancement factors optimized for solid-state applications often do not transfer well to atomic systems. This limitation has been substantiated in one of our recent works, where we explicitly showed the inadequacy of such enhancement factors for atomic KED \cite{Priya_doi:10.1080/00268976.2022.2136114}.
 %\red{This approach extends beyond traditional semi-local approximations, including generalized gradient approximations (GGAs) and meta-GGAs, by incorporating longer-range density information essential for accurately modeling metallic systems \cite{luo2018, wesolowski1997kohn, WesolowskiWang2013, wesolowski1993frozen, wesolowski1997density}. However, the electron density in atoms and molecules is far more inhomogeneous compared to that in bulk materials. Consequently, the enhancement factors or kernel functions used in non-local and GGA-type functionals optimized for solid-state applications often do not transfer well to atomic systems. This limitation has been substantiated in one of our recent works, where we explicitly showed the inadequacy of such enhancement schemes for atomic KED. \cite{Priya_doi:10.1080/00268976.2022.2136114}}.

%Stating  Alonso and Girifalco, B.M Deb and Chattaraj  work for  
For closed-shell atoms, a nonlocal approximation has been developed for calculating the KE of inhomogeneous electron systems. This approach establishes an explicit connection between the KE and the exchange energy using a correlation factor related to the electron pair distribution function \cite{Alonso1978}. By avoiding reliance on finite-order gradient expansions, it provides a more general and robust framework. In such a framework, the correlational function  
\begin{equation}
    C(\br,\br') = \frac{\rho_{2}(\br,\br')}{\rho(\br) \rho(\br')}-1
\end{equation}
is used as the starting point for further approximations. It enables the development of improved KE functionals that preserve the exact features of $T_{vW}$ while systematically enhancing the $T_{TF}$ component. Such consideration leads to a KE 
\begin{equation}\label{t_debghosh}
    T[\rho] = T_{vW}[\rho] + C_{TF}\int f(\br) \rho^{5/3}(\br) d\br.
\end{equation}
 The enhancement factor $f(\br)$ is an unknown function of $\rho(\br)$.  In the case of a homogeneous electron gas (\textit{i.e.}$\nabla \rho(\br) = 0$), the total KE reduces to $T_{TF}[\rho]$ implying $f(\br)=1$ in Eq.\eqref{t_debghosh}. Conversely, for a single-orbital (such as 1$s$-like) or bosonic systems at its ground state, $f(\br)=0$, resulting the KE to be entirely described by $T_{vW}[\rho]$. Clearly, $f(\br)$ reflects the influence of the Pauli exclusion principle and the effects of density inhomogeneity on the KE. It has also been shown to be an essential element to capture proper atomic shell structure \cite{debghosh1983}. Considerable efforts have been devoted to developing accurate, although approximate, forms of $f(\br)$ that satisfy exact theoretical constraints ~\cite{Francisco2021, Constantin2019, Priya_doi:10.1080/00268976.2022.2136114}. This admixture form of KE (Eq. \eqref{t_debghosh}) has also been employed to formulate a unified framework that combines the dynamics of quantum fluids with DFT \cite{debghosh1982_jcp, deb1989, AKRoy1999}, the resulting Deb-Chattaraj equation has been applied to study the behavior of electron densities in atoms and molecules, both under the influence of external fields and under field-free conditions \cite{akroyrsinghdeb_1997, singh_deb_96, wadehra_deb_2006, sadhukhandeb_2014}. \\
In the Kohn--Sham framework, the Pauli kinetic energy $T_{p}[\rho]$  \cite{march1960electronic, march1986local, finzelmolecule, finzel2015shell, Finzel2020, finzel2015simple, Finzel2016, finzel2016approximating, ludena2018} is the difference between Kohn-Sham KE $T_s[\rho]$ and $T_{vW}[\rho]$: $$T_{P}[\rho] = T_{s}[\rho] - T_{vW}[\rho].$$
In this setup, we can recognize $f(\br)$ as containing the similar information that is captured by the electron localization function (ELF)\cite{Becke1990}. ELF provides a bounded, inverted measure of localization and varies within the closed interval $\left[0, 1\right]$ and has been employed in diverse physico-chemical situations \cite{savin_solidstate_elf_92, ayers_elf_05, fuentealba2007_elf, nguyen25_elf}. However, the precise identification, or even a mathematically rigorous mapping of $f(\br)$ with ELF, to the best of the authors' knowledge, has not been reported so far but may be an interesting future direction. The former is defined using true KE $T[\rho]$ and the latter with Kohn-Sham KE $T_{s}[\rho]$. Unless the restriction that $T[\rho] = T_{s}[\rho]$ is put forward, any direct connection between $f(\br)$ and ELF may not be achieved.

One notable method introduced an integral formulation of density-functional theory that pioneers the use of the Suzuki-Trotter decomposition for computing many-body Green's function \cite{Weitao_yang_Prl_1987}. While this method showed promise towards a systematic development of the KED, it did not deliver the final promise that does not require orbital pictures in density functional theory.

%stating our approach
From the above discussions, it is now clear that there exists a dilemma in choosing the leading-order term for constructing the KEDs for atoms and molecules. In this backdrop, we have developed an analytical Green's function formalism (GFF), which offers a generalized and systematic framework for an accurate approximation of KED for atomic and molecular systems. In this article, we approach this very central dilemma and have shown that for realistic systems, the Thomas-Fermi KE is a fundamentally wrong leading contribution. In this approach, Green's function (GF) is the central quantity and is defined as the Laplace transform of the one-particle density matrix (ODM)
\begin{equation}\label{gfdef}
    \gfrrp = \int_{0}^{\infty} \text{d}\ef e^{-\beta \ef} \dm(\br,\br',{\ef}) \equiv \bra{\br}e^{-\beta \ham }\ket{\br'}.
 \end{equation}
Here $\ef$ is the Fermi energy and $\ham$  is the Hamiltonian of the system. The auxiliary variable $\beta$ is mathematically similar to the inverse temperature used in Statistical mechanics. However, in this formalism, the similarity is only notational and does not signify any physical thermodynamic temperature. The present formalism is a $0$ K theory. The ODM can be retrieved from the GF via inverse Laplace transform \cite{ParrYang1994}. 
\begin{equation}\label{bromwich}
  \dm(\br,\br',{\ef}) = \lim_{\Gamma \to \infty}\frac{1}{2\pi i} \int_{\gamma-i\Gamma}^{\gamma+i\Gamma}\frac{\text{d}\beta}{\beta}e^{\beta \ef} \gfrrp
 \end{equation}
 where $\gamma \in \mathbb{R}_{>} $.
 In contrast to approaches that rely on asymptotic limits, GFF takes a more general route by partitioning the total Hamiltonian according to different interactions present in the system.

Following a brief overview of existing methodologies in Section \ref{intro}, Section \ref{Theory} introduces analytical formalism developed in our recent work \cite{Priya_doi:10.1080/00268976.2022.2136114}. The formalism is then applied to a model system in Section \ref{results} to show that the constant inter-electronic repulsion is inadequate for realistic atomic systems. This article concludes with Section \ref{conclude}. 

\section{Analytical formalism} \label{Theory}
There exist previous attempts to construct a KED functional in a systematic manner using generalized gradient approximations\cite{yang1986gradient, Brack1976}. In that approach, the leading-order term corresponds predominantly to the homogeneous electron gas limit. Contrastingly, in our present approach, we partitioned our Green's function based on the relative strengths of the underlying Hamiltonians \cite{Priya_doi:10.1080/00268976.2022.2136114, priya2025new}. This approach is motivated by the perturbation expansion and explained next.
\subsection{Partition of Green's functions according to Hamiltonian strength}\label{gff}
 We begin by noting that the generic atomic Hamiltonian 
  \begin{equation}\label{ham_partition}
    \ham = 
    \underbrace{\ham_Z + \ham_0}_{\ham_H} + 
    \hat W \mathllap{\phantom{{}+\ham_0}\overbrace{\phantom{\ham_0 + \hat W}}^{ \ham_{\text{HEG}}}}
\end{equation}
 naturally partitions into the free-particle Hamiltonian $\ham_0$, the atomic charge ($Z$)-dependent nucleus-electron attraction potential $\ham_{Z}$, and interelectronic interaction potential $\hat W$. The sum of $\ham_0$ and $\ham_{Z}$ describes a hydrogenic system with Hamiltonian $\ham_{H} = \ham_0+\ham_{Z}$. The corresponding GFs are   
\begin{equation}\label{ghdef}
 G^{H}(\br, \br'; \beta) = \bra{\br}e^{-\beta \ham_{H}}\ket{\br'}
\end{equation}
and
\begin{equation}
 G^{0}(\br, \br'; \beta) = \bra{\br}e^{-\beta \ham_{0}}\ket{\br'},
\end{equation}
 respectively. For atomic systems, $\ham_{H}$ overpowers $\hat{W}$ \footnote{Consider Helium and Lithium atoms with exact non-relativistic energies as $\approx -2.904$ \textit{a.u.} and $\approx -14.864$ \textit{a.u.}, respectively. To them the hydrogenic parts of the Hamiltonian attribute $\approx -4.0$ \textit{a.u.} and $\approx -20.25$ \textit{a.u.} The contributions of the inter-electronic repulsion are evidently smaller compared to the hydrogenic part.}. This observation dictates our choice for considering $G^{H}(\br, \br'; \beta)$ to be the natural leading order term of $G(\br, \br'; \beta)$. Note that an atomic system can be approximated qualitatively better by $\ham_{Z}$ compared to the HEG limit Hamiltonian $\ham_{HEG} = \ham_0+\hat W$. Here we intend to highlight that ( see Section \ref{intro}) the approximation of the true KE of an atom has been already considered to be an admixture of two extreme limits viz. HEG ($T_{TF}$), and von Weizs\"acker ($T_{vW}$) kinetic energies.  The similar balance of these two extreme limits also manifests in information-theoretic interpretation of $T[\rho]$ \cite{Sears_1980, Romera_t_vonweizascker, gadre1991bounds}. The Fisher information, aligning with the full von Weizsäcker functional, serves as a local descriptor, and the Shannon entropy provides a complementary global view of electron delocalization \cite{A_nagy2014, SEN2007286}. Curiously, in the ``pilot wave" formulation of quantum mechanics,  the spatial average of the Bohm ``quantum potential'' \cite{bohm1952suggested,bohm1987ontological,garcia2025operator} identifies with the full von Weizs\"acker kinetic energy expression \cite{michta2015quantum, bohorquez2008local}. Such connections, although they provide at best a circumstantial basis, strengthen the hunch for the innate correctness of using an unadulterated $T_{vW}$ as the leading term of $T[\rho]$.\\
 Note that the hydrogenic Green's function $(G^{H})$ employed in our approach (Eq.\eqref{ghdef}) is of a Fermionic nature and therefore yields KE (say $T_{H}$) which is different from the $T_{vW}$. In contrast, for bosonic systems, all particles can occupy a single quantum energy state. The KE of such systems is exactly described by  $T_{vW}$. Therefore, one can qualitatively argue that both exchange-correlation hole and Pauli kinetic energy spring from the single source \emph{viz.} Pauli's exclusion principle. \cite{dyson_lenard_67,lioeb_thirring_75}. While one is not the cause or effect of the other, they have been shown to be connected \emph{via} the correlation factor (\emph{cf.} Eq.[39] and Eq. [23] of Ref. \cite{Alonso1978}, for details). Note, however, that such a connection could only be established for closed-shell systems. Intuitively, one can imagine a situation where the removal of the fermionic nature of the electrons (\emph{completely hypothetical situation}) causes all electrons to collapse into a single energy state. 
 This ``Gedankenexperiment" is thematically motivated by adiabatic switching perturbation theory \cite{weinberg_17_adpt}.
 It should be emphasized here that such ``visual construction'' does not represent the actual situation completely but provides a useful ``classical'' conception of Pauli KE.  This ``removal of fermion-ness'' simultaneously ``eliminates'' both the exchange hole and the Pauli KE. Conceptually, such a ``process'' would transform $T_{H}$ into the von Weizsäcker KE $T_{vW}$, the latter being the KE of the bosonic system. This outcome is consistent with the previously established,  mathematically rigorous result that $T_{H}> T_{vW}$ \cite{HoffmannOstenhof_1977, Lieb_1983,gadre1991bounds}, bolstering our qualitative picture.\\
 %**This was the original paragraph
 %Consider a Bosonic system where all the population is on the ground state. On the other hand, the Fermions follow the Pauli exclusion principle leading to exchange-hole. The exchange-hole is responsible for pushing the electrons to higher energy states\cite{dyson_lenard_67,lioeb_thirring_75}. If we switch off the exchange-hole, all electrons occupy the ground state $1s$. As a result, $T_{H}$ transforms naturally $T_{vW}$. Since the KE of the excited states is higher than that of the ground state, it can be argued that $T_{H}> T_{vW}$ \red{ \cite{HoffmannOstenhof_1977, Lieb_1983,gadre1991bounds}}. 
 %***
 The interaction Green's function $(G^{int})$ is defined as the difference between total Green's function $(G)$ and the hydrogenic Green's function $(G^{H})$ 
 \begin{equation}\label{g_rrp_rep}
   G^{int}(\br, \br'; \beta) \equiv \bra{\br}\hat O\ket{\br'} = G(\br, \br'; \beta)- G^{H}(\br, \br'; \beta).
 \end{equation}
 Here 
 \begin{equation}
     \hat O = e^{-\beta \ham} - e^{-\beta \ham_{H}}.
 \end{equation}
%Kuch? Bhi@2411
The central challenge for our formalism is to construct a decent approximate expression for $G^{int}$ which describes a universal part of the full GF having no system-dependent term in it. 
  In order to get the systematic expansion of $G^{int}$ we expand $\hat{O}$ using Zassenhaus formula \cite{CasasNadinic2012, magnus1954} as
 \begin{equation}\label{hato_zassenhaus}
     \hat O = - e^{-\beta \ham_{H}}\left( 1-e^{-\beta \hat W} e^{-\frac{\beta^2}{2} \comm{\ham_H}{\hat W}} \ldots \right).
 \end{equation}
%We find
The fact that
\begin{equation}\label{comm_H}
    \comm{\ham_H}{\hat W}=\comm{\ham_0}{\hat W} 
\end{equation}
inspired us to approximate all higher order commutators inside the parentheses in the RHS involving $\ham_H$ to be replaced by $\ham_0$ 
%since $\comm{\ham_Z}{\hat W}$ commutes. However, such replacements cannot be done for terms containing triple commutators onwards, 
leading to our first approximation

   \begin{equation}\label{hatohydr_heg}
     \hat O \approx e^{-\beta \ham_{H}} e^{\beta \ham_{0}}\left(e^{-\beta (\ham_0+\hat W)} - e^{-\beta \ham_{0}}\right) .
\end{equation} 
Despite this approximation, the expansion of the second-order commutator represents a partial sum up to the infinite order interaction in a similar spirit of random phase approximation \cite{gellmann_bruckner_57}. Additionally, this approximation is qualitatively acceptable where $\ham_0 \approx \ham_H$,  away from the nucleus where $\ham_Z \to 0$. Note that we have not completely replaced $\ham_{Z}$ by $\ham_{0}$ since the leading term of the expansion is still $G^{H}(\br, \br', \beta)$. Also, this approximation is consistent with the fact that $f(\br) \to 1$ away from the nucleus. We found, however, that this approximation leads to an unphysical asymmetry in the resulting GF with respect to $\br$ and $\br'$. 
To minimize the effect of asymmetry, we decided to use a symmetrized GF
\begin{equation}\label{symgf}
    {G}(\br, \br';\beta) \equiv \frac{1}{2}\left[\gfrrp+\gfrpr\right]
\end{equation}
as our fundamental GF where $\gfrrp$ in the RHS of the above equation is defined as 
 \begin{equation}\label{resoid}
    \gfrrp = {G^{H}}(\br, \br'; \beta)+ \int \dbr'' \bra{\br}e^{-\beta \ham_{H}}\ket{\br''}\bra{\br''}\sum_{n=0}^{\infty} \frac{\beta^n}{n!}\ham_{0}^{n}\left(e^{-\beta (\ham_0+\hat W)} - e^{-\beta \ham_{0}}\right) \ket{\br'}.
 \end{equation}   
Eqs.\eqref{symgf} and \eqref{resoid}, taken together, are the governing equations to systematically calculate the KED.  The convergent infinite sum over $n$ in Eq.\eqref{resoid} guarantees a systematic scheme of approximations as desired. %This feature of our formalism allows for a rigorous development scheme without re-coursing to any parameter tuning. Each successive approximation will provide greater accuracy in the similar spirit of ``Jacob's ladder'' in XC functionals. 
In our scheme, as a result, no empirical parameter would be required, and further accuracy can be readily obtained by introducing higher-order terms.  
\section{Results and Discussion} \label{results}
The actual computation of KE for real atoms and molecules, via our formulation, will require the Green's function of the three-dimensional hydrogenic system as well as an accurate mean-field interelectronic repulsion potential that,in principle, contains exchange-correlation effects.  
 Before plunging into such a full-blown \emph{tour-de-force}, in this paper, we have considered a particle-density (with particle mass $m$) moving under a Pöschl-Teller potential 
 \begin{equation}
    {V}_{PT} (x)= -\frac{\lambda (\lambda+1)}{2}\sech^2(x) 
\end{equation}
 while interacting among themselves by a mean-field interelectronic repulsive force $W(x)$. The depth of the Pöschl-Teller potential is determined by the parameter $\lambda$. The bound energy eigenstates of the potential for fixed $\lambda$ are given by 
 \begin{equation}
     \Psi_{\lambda}^{\mu}(x) = N_{\lambda}^{\mu}P_{\lambda}^{\mu}(\tanh(x))
 \end{equation}
 with energy eigenvalues 
 \begin{equation}
     E_{\mu} = -\frac{\hslash^2\mu^2}{2m}.
 \end{equation}
\begin{figure}
\centering
\includegraphics[scale=0.4]{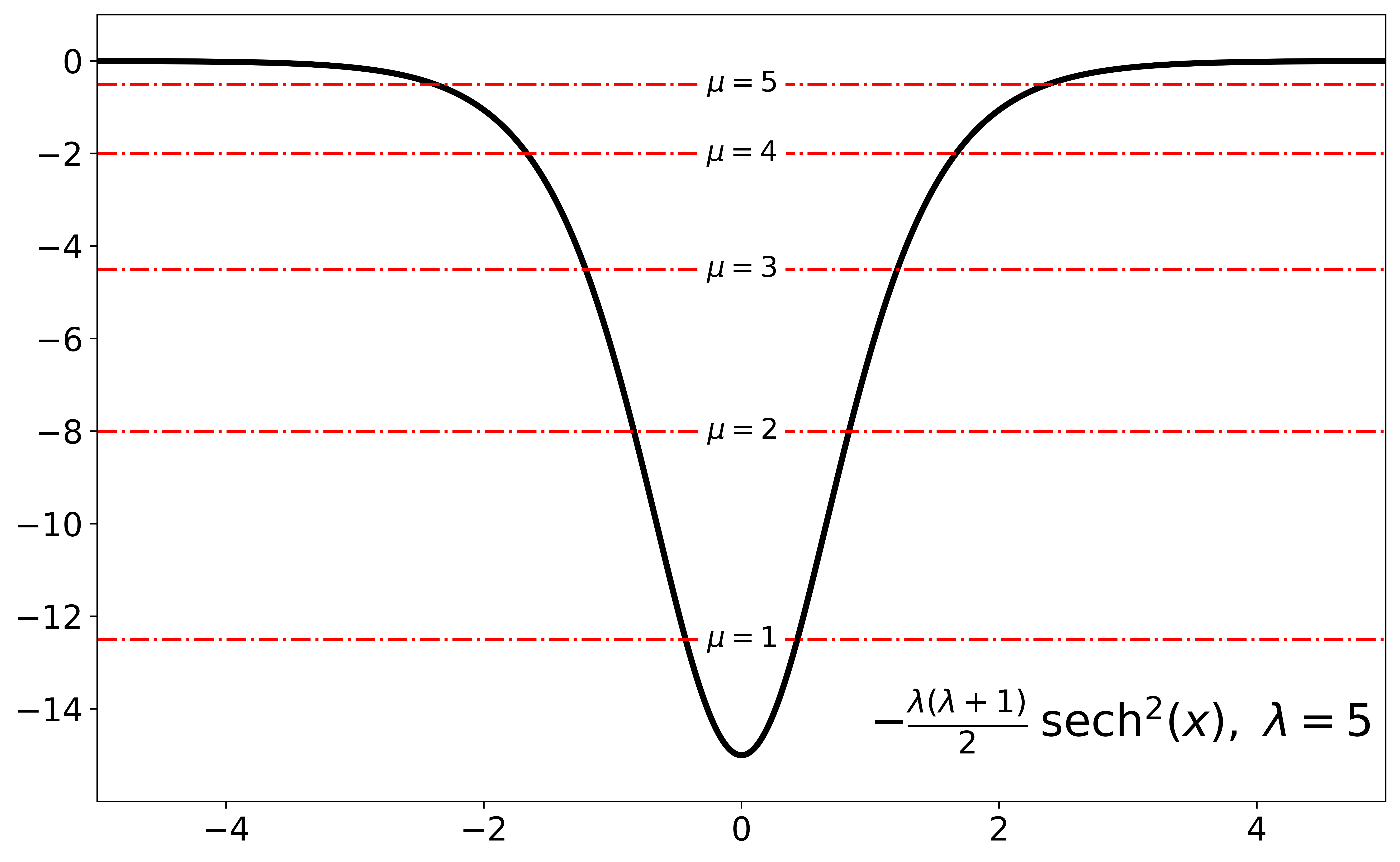}
\caption{P\"oschl-Teller potential for $\lambda=5$ having five bound states.}\label{fig:ptpoten}
\end{figure}
Here $\mu = 1, 2, \ldots, \lambda$.
The P\"oschl-Teller potential, as shown in Fig.\ref{fig:ptpoten}, sustains both bound and scattering states, thereby closely resembling the actual hydrogenic attractive nucleus-electron interaction potential. For this system, the Eq.\eqref{resoid} becomes 
    \begin{equation} \label{GFmain}
   \gfxxp = {G}^{PT}(x, x'; \beta)+ \int dx'' {G}^{PT}(x, x''; \beta) \bra{x''}\sum_{n=0}^{\infty} \frac{\beta^n}{n!}\ham_{0}^{n}\left(e^{-\beta (\ham_0+\hat W)} - e^{-\beta \ham_{0}}\right) \ket{x'}.  
\end{equation}
Our core objective of this paper is to analyze the qualitative features of the present method. To accomplish that goal, we first consider system is interacting via a weak, homogeneous interelectronic potential. 
\subsection{Weak interaction limit}\label{gfweakfieldposchlteller}
%To exemplify the efficacy of our GFF we apply it to a one-dimensional Pöschl-Teller . Also, we have refrained from using the atomic units in the following sections to clarify the relative magnitudes of different terms.  The governing equation for Pöschl-Teller is 
%
%For this system, 
%\begin{widetext}
 %   \begin{equation} \label{GFmain}
 %  \gfxxp = {G}^{PT}(x, x'; \beta)+ \int dx'' {G}^{PT}%(x, x''; \beta) \bra{x''}\sum_{n=0}^{\infty} \frac{\beta^n}{n!}\ham_{0}^{n}\left(e^{-\beta (\ham_0+\hat W)} - e^{-\beta \ham_{0}}\right) \ket{x'}.  
%\end{equation}
%\end{widetext}
%In this equation, $\hat{W}$ is the mean-field interaction potential.
We have considered the weak-field limit since it resembles the low-density limit of the Gell-Mann-Bruckner model\cite{gellmann_bruckner_57} where particles move under a constant positive background, leading to homogeneous particle density. As a result, the mean-field interaction potential also varies negligibly over space. To model this behavior, we have expanded the mean-field, one-dimensional potential $W(x)$ in a power series 
\begin{equation}\label{weakfield}
     W(x) = W_{0} + x \partial_{x} W(x)|_{0}+ \hdots 
 \end{equation}
and considered only the leading-order constant term $W_{0}$.  This approximation is equivalent to $\ham_0 \approx \ham_H$. 
%In this approach, we are trying to lay road map  on how to calculate the KE from our method. First the calculation on is done for weak interaction potential. 
By virtue of this approximation, Eq.\eqref{GFmain} becomes
%\begin{widetext}
\begin{equation}\label{gf_shofree}
 \gfxxp = {G}^{PT} (x, x' ; \beta) + \int dx'' (e^{-\beta  W_{0}}-1 ) \left(    {G}^{PT}(x, x'' ; \beta) e^{-\beta \frac{\partial}{\partial\beta}} {G}^{0}(x'', x' ; \beta) \right).
\end{equation}    
We finally obtain the KE corresponding to the free PT potential (i.e. for $W_0 = 0$) as (see  Appendix \ref{kept} for details)
\begin{equation}
    T[\rho]= -\frac{\hslash^2}{2m} \sum_{\mu=1}^{\lambda} \left( \mu^2- \frac{2\lambda  (\lambda +1)\mu}{2\lambda+1} \right) \Theta(\ef+ \frac{\hslash^2 \mu^2}{2m})
\end{equation}
where $\ef > -\frac{ \hslash^2 \mu^2}{2m} $ while
%then it is one otherwise $0$
for the KE part due to constant interaction (see Appendix \ref{keintpt} for details) is
\begin{equation}
    T_{int}[\rho] = -\frac{1}{2}\frac{\hslash^2}{2m} \sum_{\mu=1}^{\lambda} \left( \mu^2- \frac{2\lambda  (\lambda +1)\mu}{2\lambda+1} \right) \left( \Theta(\ef -W_{0}+\frac{\hslash^2 \mu^2}{2m} ) -\Theta(\ef+ \frac{\hslash^2 \mu^2}{2m}) \right).
\end{equation}

Here we wish to check the effect of $\lambda$ since $\lambda$ mimics the effect of the atomic number $Z$. For the ground state where $\lambda=\mu$, the KE for this system becomes
$T = \frac{3\hslash^2}{4m} \frac{-\lambda^2}{2\lambda+1}$ which shows that under constant repulsive interaction the KE does not follow the expected variation with respect to system size \cite{ParrYang1994}. This result reinforces our suspicion that the homogeneous electron-gas limit is not adequate for the realistic atomic system and should not be considered as the mainstay of the design of the KE of such systems. Note that for the simple harmonic oscillator, the bound state limit faithfully produces the $Z$-dependence, even for a homogeneous electron gas. Therefore, this result begs for further analysis of the Green's function for a more realistic interelectronic interaction potential.    
%%% Plot of T vs lambda and discussion over it
%%% along with comparison with SHO result
\subsection{Comparison of Green's function with Coulomb interaction limit}
\begin{figure}[H]
    \centering
    \includegraphics[width=0.95\linewidth]{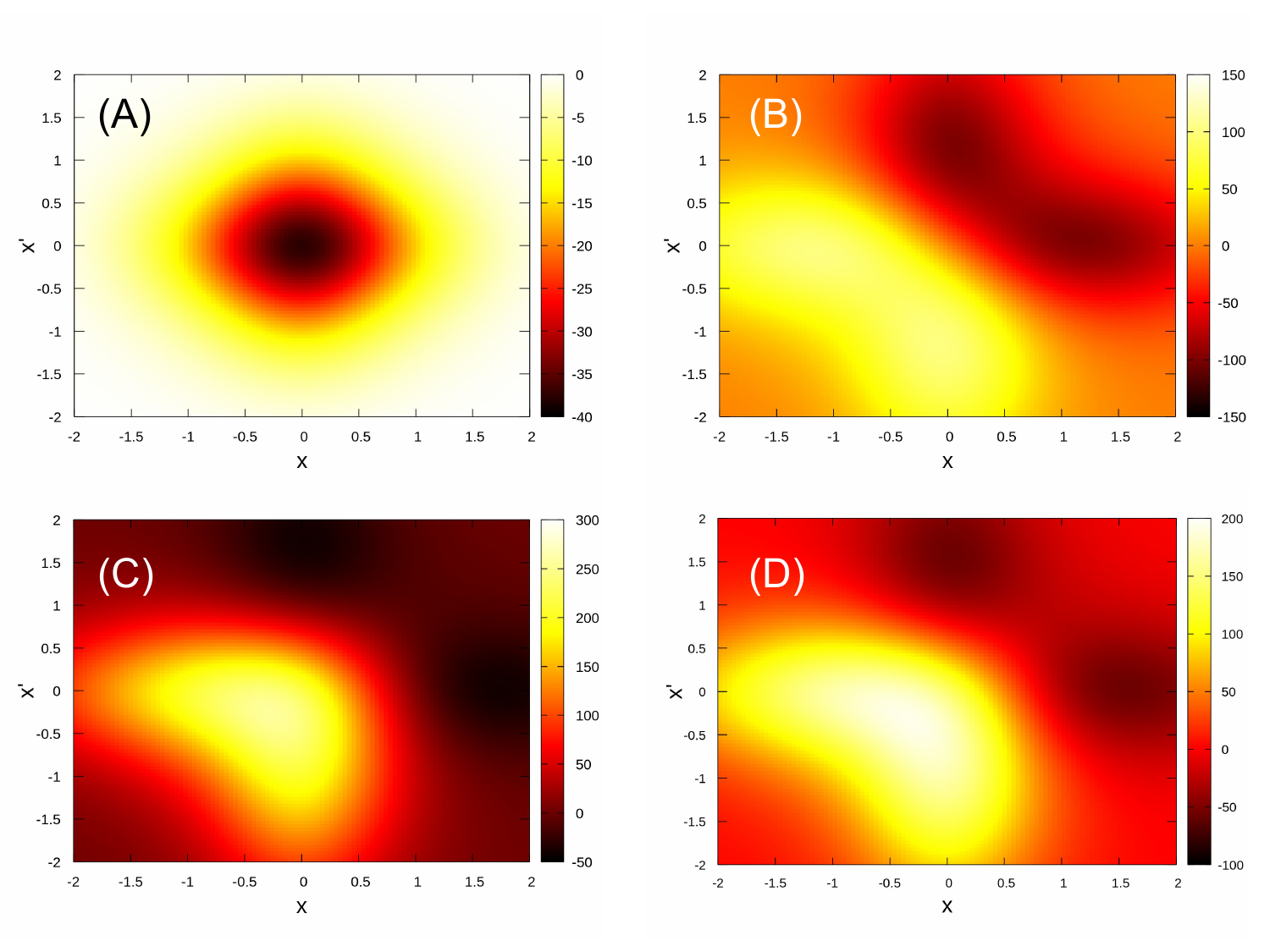}
    \caption{Contour plots of  $G^{int}(x,x';\beta)$ for constant interaction potential (A),  Yukawa potential for M=1 (B), M=2 (C) and M=3 (D). Using the     $\beta= 1.0, g= 1.0$ }
    \label{fig:gint}
\end{figure}

To compare the non-local natures of Green's function of the constant potential and a more realistic potential, we have employed a non-local, Yukawa-like potential in momentum space
\begin{equation}
    V(k, k') = -\frac{g^2}{(k-k')^2}
\end{equation}
where $g$ determines the strength of the repulsive interaction. Since $\ham_{0}$ and $\hat{W}$ do not commute in Eq.\eqref{GFmain}, we applied the Suzuki-Trotter decomposition and retained the zeroth-order term, yielding: 
\begin{equation}
   \gfxxp = {G}^{PT}(x, x'; \beta)+ \underbrace{\int dx'' {G}^{PT}(x, x''; \beta) \bra{x''}\left( \lim_{M\longrightarrow \infty} \left(e^{-\frac{\beta}{M} \ham_0 }e^{\frac{\beta}{M}\hat W} \right)^M - e^{-\beta \ham_{0}}\right) \ket{x'}}_{G^{int}(x,x';\beta)}.   
\end{equation}
Here $\bra{x''}\left( \lim_{M\longrightarrow \infty} \left(e^{-\frac{\beta}{M} \ham_0 }e^{\frac{\beta}{M}\hat W} \right)^M  - e^{-\beta \ham_{0}}\right)\ket{x'} = h_{int}(x'', x')$, is universal and system-independent. Therefore, we have solved this part separately and compared with the corresponding $G^{int}(x,x';\beta) $ term for constant $W_0$ (detailed derivation can be found in SI) in Fig.\ref{fig:gint}.
To explore the qualitative nature of the approximation, we evaluate the Suzuki–Trotter expansion for $M=1,2,3$. We can clearly see that the qualitative features for the constant inter-electronic repulsion (Fig.\ref{fig:gint}(A)) looks very different compared to the $M=1$ (Fig.\ref{fig:gint}(B)), $M=2$ (Fig.\ref{fig:gint}(C)) and $M=3$ (Fig.\ref{fig:gint}(D)). Also, we see the $M =2$ and $M=3$ look more similar compared to $M =1$ case, assuring a quick convergence of Suzuki-Trotter terms. These contour plots unquestionably prove that the homogeneous electron gas model differs qualitatively as well as quantitatively from the atomic systems where the bound states dominate the energy spectrum. In the near future, we will employ this knowledge to construct a better approximation of the KED of atoms. 
%\subsection{ Spatial nature of Green's function for coulomb inetraction in P\"oschl-Teller}
%\begin{figure}
 %   \centering
 %   \includegraphics[width=0.5\linewidth]{fig_jcc/fit_tbetamasked4_betan.png}
  %  \caption{Rough fitted plot for $t(\beta)$ Parameter $g=5$ for m=1, fitted function $ (f(\beta) = A \beta^n e^{-b \beta})$,$A= 3.2291, n=-0.3013$ and $b=0.7941$  }
   % \label{fig:enter-labe}
%\end{figure}

%\begin{figure}
%    \centering
%    \includegraphics[width=0.5\linewidth]{fig_jcc/gintcontour_beta5_g0p5.png}
%    \caption{The parameter $\beta= 5, g= 0.5$}
 %   \label{fig:enter-label}
%\end{figure}
% need to write it properly
%\subsection{Coulomb interaction limit for simple  harmonic oscillator}

\section{Conclusion} \label{conclude}
In summary, we have shown that the homogeneous electron gas limit is a fundamentally flawed starting point for the development of an atomic kinetic energy density. This conclusion is drawn from modeling the nuclear attraction by a P\"oschl-Teller potential. It has been argued that the P\"oschl-Teller potential provides a more realistic model for atomic potential compared to the simple harmonic oscillator model. In contrast to the later model, where a constant electron density limit produces a qualitatively acceptable $N$-dependence, our present work emphasizes the need for essentially non-uniform density-dependent inter-electronic potential for realistic systems. As a consequence, this work settles the issue of choosing the leading order term for an atomic KED in favor of von Weizs\"acker KED. We also have furthered our previous work with Green's function formulation by introducing detailed techniques to compute kinetic energy density for realistic systems. The actual computation of the local and global kinetic energy density of atoms and molecules using our formalism is a natural way forward and requires further work. For such computations, it is expected that the qualitative conclusions of this work will be retained, while the actual KE values will obviously change to actual numerical values of KE. This statement has been motivated by recent developments due to Theophilou's theorem and related works \cite{theophilou_2018, nagy_dft_2024} which ensures a one-to-one mapping between the radially averaged electron density around a nucleus and the external potential for arbitrary Coulombic systems. The only point that requires further investigation is the qualitative effect of singularity in the actual Coulomb potential on the kinetic energy. Since such a singularity structure is absent in the P\"os chl-Teller potential, we will refrain from speculating on its effects further.  In essence, this article highlights yet another rationale for the adoption of the Hamiltonian partitioning scheme in conjunction with the Green's function technique for an accurate and systematic method for the kinetic energy density for atoms and molecules.

\backmatter

\bmhead{Supplementary information}
Supplementary Information is provided for clarity and better comprehension.

\bmhead{Acknowledgements}
Priya acknowledges a Prime Minister's Research Fellowship and MS acknowledges IIT Kanpur initiation grant no. IITK/CHM/2018419 and SERB startup research grant no.
SRG/2019/000369 for partial supports.  All authors are deeply indebted to Ms. Sangini Gupta and Mr. Abhinav Aryan for their friendship and enthusiasm.

\section*{Declarations}
\begin{itemize}
\item Funding:\\
Priya acknowledges a Prime Minister's Research Fellowship. MS acknowledges IIT Kanpur initiation grant no. IITK/CHM/2018419 and SERB startup research grant no.
SRG/2019/000369 for partial supports. 
\end{itemize}

\begin{appendix}    
\section{Detailed derivation of weak interaction limit for  Pöschl-Teller potential} \label{kept}
\begin{equation}\label{tsdef}
     T[\rho] = \int t(x, \rho(x)) dx
 \end{equation}
 where
\begin{equation}
    t(x, \rho(x)) =-\frac{\hslash^2}{2m} \left(\frac{d^2}{dx^2}\rho_{1}(x,x';\beta)\right)\vert_{x=x'}.
\end{equation}
\begin{eqnarray}
    T[\rho]&=& -\frac{\hslash^2}{2m} \int {\frac{\partial^2}{\partial x^2} (\rho_{1}(x,x'; \beta))}\vert_{x=x'} dx
\end{eqnarray}
\begin{eqnarray}
    T[\rho]&=& -\frac{\hslash^2}{2m} \int {\frac{\partial^2}{\partial x^2}} \left( \lim_{T\to \infty}\frac{1}{2\pi i} \int_{\gamma-iT}^{\gamma+iT}\frac{\text{d}\beta}{\beta}e^{\beta \ef} \gfxxp \right) \vert_{x=x'} dx \notag\\
    &=& -\frac{\hslash^2}{2m} \lim_{T\to \infty}\frac{1}{2\pi i} \int_{\gamma-iT}^{\gamma+iT}\frac{\text{d}\beta}{\beta}e^{\beta \ef} t(\beta)  
\end{eqnarray}
Where, we have used $t(\beta)= \int {\frac{\partial^2}{\partial x^2}}\gfxxp \vert_{x=x'}$ dx. \\
In order to find the kinetic energy for the P\"oschl-Teller, we did $\frac{d^2}{dx^2}$
\begin{equation}\label{kerneld2x}
    \frac{d^2}{dx^2}\bra{x} e^{-\beta \hat{H}_{PT}} \ket{x'} =   \sum_{\mu=1}^{\lambda} (N_{\lambda}^{\mu})^2 \frac{d^2}{dx^2} P_{\lambda}^{\mu}(\tanh{x}) P_{\lambda}^{\mu}(\tanh{x'}) e^{\frac{\beta \hslash^2 \mu^2}{2m}}
\end{equation}
To find the second derivative of the associated Legendre function \( P_{\lambda}^{\mu}(\tanh(x)) \) with respect to \( x \). It can be done by applying the chain rule and product rule for differentiation.\\ 
First, we start by denoting \( y = \tanh(x) \) and using the associated Legendre differential equation:
%The first derivative of \( P_{\lambda}^{\mu}(y) \) with respect to \( x \) is:
%\begin{equation}
    %\frac{d}{dx} P_{\lambda}^{\mu}(y) = \frac{d}{dy} P_{\lambda}^{\mu}(y) %\cdot \frac{dy}{dx} = P_{\lambda}^{\mu'}(y) \cdot %\operatorname{sech}^2(x)
%\end{equation}

%Next, we compute the second derivative using the product rule:
%\begin{equation}
   % \frac{d^2}{dx^2} P_{\lambda}^{\mu}(y) = \frac{d}{dx} \left[ P_{\lambda}^{\mu'}(y) \cdot \operatorname{sech}^2(x) \right]
%\end{equation}

%Applying the chain rule to the first term and the derivative of \( \operatorname{sech}^2(x) \) to the second term, we get:
%\begin{equation}
   % \frac{d}{dx} \left[ P_{\lambda}^{\mu'}(y) \right] \cdot \operatorname{sech}^2(x) + P_{\lambda}^{\mu'}(y) \cdot \frac{d}{dx} \left[ \operatorname{sech}^2(x) \right]
%\end{equation}

%The derivative of \( P_{\lambda}^{\mu'}(y) \) with respect to \( x \) is:
%\begin{equation}
    %\frac{d}{dx} P_{\lambda}^{\mu'}(y) = P_{\lambda}^{\mu''}(y) \cdot \operatorname{sech}^2(x)
%\end{equation}

%The derivative of \( \operatorname{sech}^2(x) \) is:
%\begin{equation}
    %\frac{d}{dx} \operatorname{sech}^2(x) = -2 \operatorname{sech}^2(x) \tanh(x)
%\end{equation}

%Substituting these results back, we have:
%\begin{equation}
    %P_{\lambda}^{\mu''}(y) \cdot \operatorname{sech}^4(x) - 2 P_{\lambda}^{\mu'}(y) \cdot \operatorname{sech}^2(x) \tanh(x)
%\end{equation}
\begin{equation}
    (1 - y^2) P_{\lambda}^{\mu''}(y) - 2y P_{\lambda}^{\mu'}(y) + \left[ \lambda(\lambda + 1) - \frac{\mu^2}{1 - y^2} \right] P_{\lambda}^{\mu}(y) = 0
\end{equation}
Solving for \( P_{\lambda}^{\mu''}(y) \):
\begin{equation}
    P_{\lambda}^{\mu''}(y) = \frac{2y P_{\lambda}^{\mu'}(y) - \left[ \lambda(\lambda + 1) - \frac{\mu^2}{1 - y^2} \right] P_{\lambda}^{\mu}(y)}{1 - y^2}
\end{equation}
Substituting \( y = \tanh(x) \) and \( 1 - y^2 = \operatorname{sech}^2(x) \), we simplify the expression. After substituting and simplifying, we find that the second derivative is:
\begin{equation}
    \left[ \mu^2 - \lambda(\lambda + 1) \operatorname{sech}^2(x) \right] P_{\lambda}^{\mu}(\tanh(x))
\end{equation}
Thus, the final answer is:
\begin{equation}\label{d2alp}
    \frac{d^2}{dx^2} P_{\lambda}^{\mu}(y)=\left[ \mu^2 - \lambda(\lambda + 1) \operatorname{sech}^2(x) \right] P_{\lambda}^{\mu}(\tanh(x))
\end{equation}
Now substituting the Eq.\eqref{d2alp} in Eq.\eqref{kerneld2x} and substituting $x=x'$ then integration with respect to $x$, the resulting equation is as follows 
\begin{equation}
     \int_{-\infty}^{\infty} dx  \sum_{\mu=1}^{\lambda} (N_{\lambda}^{\mu})^2 \left[ \mu^2 - \lambda(\lambda + 1) \operatorname{sech}^2(x) \right] P_{\lambda}^{\mu}(\tanh(x))^2 e^{\frac{\beta \hslash^2 \mu^2}{2m}}
\end{equation}
%Dropping $\sum_{\mu=1}^{\lambda} (N_{\lambda}^{\mu})^2 e^{\frac{\beta \hslash^2 \mu^2}{2m}}$ for simplicity and focusing on the integration.
\begin{equation}
    \sum_{\mu=1}^{\lambda} (N_{\lambda}^{\mu})^2 e^{\frac{\beta \hslash^2 \mu^2}{2m}} \int_{-\infty}^{\infty} dx  \left[ \mu^2 - \lambda(\lambda + 1) \operatorname{sech}^2(x) \right] P_{\lambda}^{\mu}(\tanh(x))^2 %e^{\frac{\beta \hslash^2 \mu^2}{2m}}
\end{equation}
To integrate w.r.t x, we consider $y= \tanh{x}$ and $1-y^2=\operatorname{sech}^2(x)$; therefore, the integral variable is changed from $x$ to $ y$, and the range becomes $[-1,1]$.
\begin{equation}
    \int_{-1}^{1} \left[\mu^2 - \lambda (\lambda +1)(1 - y^2) \right] \frac{P_{\lambda}^{\mu}(y)^2}{1 - y^2}  dy
\end{equation}

\begin{equation}
   \mu^2 \int_{-1}^{1}  \frac{P_{\lambda}^{\mu}(y)^2}{1 - y^2}  dy - \lambda (\lambda +1) \int_{-1}^{1}P_{\lambda}^{\mu}(y)^2 dy
\end{equation}
For solving this integral, we use orthogonality identities for associated Legendre polynomials:
\begin{equation}
    \int_{-1}^{1} P_{\lambda}^{\mu}(x) P_{\lambda'}^{\mu}(x) \, dx = 
    \frac{2}{2\lambda+1} \frac{(\lambda+\mu)!}{(\lambda-\mu)!} \delta_{\lambda \lambda'}
\end{equation}
\begin{equation}
\int_{-1}^{1} P_{\lambda}^{\mu} P_{\lambda}^{\nu} \frac{dx}{1 - x^2} =
\begin{cases} 
0, & \text{if }\mu \neq \nu \\[8pt]
\frac{(\lambda + \mu)!}{\mu(\lambda - \mu)!}, & \text{if } \mu = \nu \neq 0 \\[8pt]
\infty, & \text{if } \mu = \nu = 0
\end{cases}
\end{equation}
Using these identities, we can find the normalization constant also
\begin{align}
    &(N_{\lambda}^{\mu})^2 \int_{-1}^{1}  \frac{P_{\lambda}^{\mu}(y)^2}{1 - y^2}dy=1 \notag\\
    \implies&(N_{\lambda}^{\mu})^2= \frac{\mu (\lambda -\mu)!}{(\lambda +\mu)!}
\end{align}
Therefore, the final form of the integral is 
\begin{equation}
  \frac{\mu(\lambda + \mu)!}{(\lambda - \mu)!}- \frac{2\lambda (\lambda +1)}{2\lambda+1} \frac{(\lambda+\mu)!}{(\lambda-\mu)!} 
\end{equation}
and the complete expression of $t(\beta)$ can be written as 
\begin{align} \label{ke_int}
    &\sum_{\mu=1}^{\lambda} (N_{\lambda}^{\mu})^2\left( \frac{\mu(\lambda + \mu)!}{(\lambda - \mu)!}- \frac{2\lambda (\lambda +1)}{2\lambda+1} \frac{(\lambda+\mu)!}{(\lambda-\mu)!} \right) e^{\frac{\beta \hslash^2 \mu^2}{2m}} \notag\\ 
    -&\frac{\hslash^2}{2m} \lim_{T\to \infty}\frac{1}{2\pi i} \int_{\gamma-iT}^{\gamma+iT}\frac{\text{d}\beta}{\beta}e^{\beta \ef}\sum_{\mu=1}^{\lambda} (N_{\lambda}^{\mu})^2\left( \frac{\mu(\lambda + \mu)!}{(\lambda - \mu)!}- \frac{2\lambda (\lambda +1)}{2\lambda+1} \frac{(\lambda+\mu)!}{(\lambda-\mu)!} \right) e^{\frac{\beta \hslash^2 \mu^2}{2m}}
\end{align}    
The resultant kinetic energy contribution turns out to be, on condition that $\ef > -\frac{ \hslash^2 \mu^2}{2m} $, then it is one, otherwise $0$.

\begin{equation}
    T[\rho]= -\frac{\hslash^2}{2m} \sum_{\mu=1}^{\lambda} \left( \mu^2- \frac{2\lambda  (\lambda +1)\mu}{2\lambda+1} \right) \Theta(\ef+ \frac{\hslash^2 \mu^2}{2m})
\end{equation}

\section{Calculation for the interaction part of the Pöschl-Teller potential for constant interaction potential ($W_{0}$)}\label{keintpt}
\begin{equation} 
   \gfxxp = {G}^{PT}(x, x'; \beta)+ \int dx'' {G}^{PT}(x, x''; \beta) \bra{x''}\sum_{n=0}^{\infty} \frac{\beta^n}{n!}\ham_{0}^{n}\left(e^{-\beta (\ham_0+\hat W)} - e^{-\beta \ham_{0}}\right) \ket{x'}.  
\end{equation}
The basic definition used in our work:
\begin{equation}
      {G}^{PT}(x, x''; \beta)  =  \sum_{\mu=1}^{\lambda}(N_{\lambda}^{\mu})^2 P_{\lambda}^{\mu}(\tanh{x}) P_{\lambda}^{\mu}(\tanh{x''}) e^{\frac{\beta \hslash^2 \mu^2}{2m}}
  \end{equation}
%where $N_{\lambda}^{\mu}$ is the normalization constant and $P_{\lambda}^{\mu}(\tanh x)$ is the associated Legendre functions and $E_{\mu} =-\frac{\hslash^2 \mu^2}{2m}$ are the eigenvalues of the Hamiltonian corresponding to the P\"oschl-Teller potential.\\  
The ${G}^{0}(x, x''; \beta) $ is  as follows:
  \begin{align}
     &{G}^{0}(x'', x'; \beta) = A_{0} \exp{-B_{0} (x''-x')^{2}}, \notag\\ \text{ where }
     &A_{0} = \left(\sqrt{\frac{m}{2\pi \beta \hslash ^{2}}}\right), B_{0}=\left(\frac{m}{2 \beta \hslash^{2}}\right).  
\end{align}
%\begin{eqnarray*}
   %A_{0}&=&\sqrt{\frac{m}{2\pi \beta \hslash ^{2}}} \\
    %B_{0}&=& \frac{m}{2 \beta \hslash^{2}}. 
%\end{eqnarray*}
We considered the weak interaction limit 
\begin{equation}
     W(x) = W_{0} + x \partial_{x} W(x)|_{x=0}+ \hdots 
 \end{equation}
We have taken the interaction potential to be constant, as a result
eq. \eqref{GFmain} becomes
 \begin{equation}
     \gfxxp = {G}^{PT}(x, x'; \beta)+ \int dx'' {G}^{PT}(x, x''; \beta) \bra{x''}\sum_{n=0}^{\infty} \frac{\beta^n}{n!}\ham_{0}^{n}\left(e^{-\beta \ham_0}e^{-\beta \hat W_{0}} - e^{-\beta \ham_{0}}\right) \ket{x'}.  
 \end{equation}
 Rewriting the above equation
 \begin{align}
     &\gfxxp = {G}^{PT}(x, x'; \beta)+ \int dx'' {G}^{PT}(x, x''; \beta) (e^{-\beta W_{0}}-1)\bra{x''}\sum_{n=0}^{\infty} \frac{\beta^n}{n!}\underbrace{ \ham_{0}^{n} e^{-\beta \ham_{0}}}_{\frac{\partial^{n}}{\partial(-\beta)^{n}} e^{-\beta \ham_{0}}} \ket{x'} \notag\\ 
     \implies&\gfxxp = {G}^{PT}(x, x'; \beta)+ (e^{-\beta W_{0}}-1)\int dx'' {G}^{PT}(x, x''; \beta) e^{-\beta \frac{\partial}{\partial\beta}} G^{0}  (x'',x'; \beta)
\end{align}
Solving for the 2nd part, we are approximating $e^{-\beta \frac{\partial}{\partial\beta} }G^{0}  (x'',x'; \beta) \approx \left (1-\beta \frac{\partial}{\partial \beta} \right)G^{0}  (x'',x'; \beta)= \frac{3}{2}G^{0}  (x'',x'; \beta)- B_{0} (x''-x')^2 G^{0}  (x'',x'; \beta)$.\\
Therefore, we get, 
\begin{eqnarray}\label{main1D}
  \gfxxp &=& {G}^{PT}(x, x'; \beta)+ (e^{-\beta W_{0}}-1) \left(\frac{3}{2}\int dx''{G}^{PT}(x, x''; \beta)G^{0}  (x'',x'; \beta)\right)\\
  &-& (e^{-\beta W_{0}}-1)\left(B_{0}\int dx''{G}^{PT}(x, x''; \beta) (x''-x')^2 G^{0}  (x'',x'; \beta) \right)  
\end{eqnarray}
Solving the second term, taking the integral parts only 
\begin{align}
    I_{2} &= \int dx''{G}^{PT}(x, x''; \beta)G^{0}  (x'',x'; \beta) \\
   I_{2} &= A_{0}\sum_{\mu=1}^{\lambda}(N_{\lambda}^{\mu})^2 e^{\frac{\beta \hslash^2 \mu^2}{2m}}\int  P_{\lambda}^{\mu}(\tanh{x}) P_{\lambda}^{\mu}(\tanh{x''})  e^{-B_{0}(x''-x')^2} dx''
\end{align}
For analytical solution expanding $P_{\lambda}^{\mu}$ around $x''=x'$, 
\begin{equation}
    P_{\lambda}^{\mu}(\tanh{x''}) = \sum_{k=0}^{\infty}\frac{(x''-x')^k}{k!} \frac{d^k}{dx^k}P_{\lambda}^{\mu}(\tanh(x'')\vert_{(x''=x')}
\end{equation}
Considering the first term only as the second term is odd, the integral will be zero. The analytical expression to solve becomes 
\begin{eqnarray}
    I_{2} &=& A_{0}\sum_{\mu=1}^{\lambda}(N_{\lambda}^{\mu})^2 \int  P_{\lambda}^{\mu}(\tanh{x}) P_{\lambda}^{\mu}(\tanh{x'}) e^{\frac{\beta \hslash^2 \mu^2}{2m}}e^{B_{0}(x''-x')^2} dx'' \notag\\ 
    I_{2} &=& A_{0} \sum_{\mu=1}^{\lambda}(N_{\lambda}^{\mu})^2 \sqrt{\frac{\pi}{B_{0}}} P_{\lambda}^{\mu}(\tanh{x}) P_{\lambda}^{\mu}(\tanh{x'}) e^{\frac{\beta \hslash^2 \mu^2}{2m}} \notag\\
    &=&  \sum_{\mu=1}^{\lambda}(N_{\lambda}^{\mu})^2 P_{\lambda}^{\mu}(\tanh{x}) P_{\lambda}^{\mu}(\tanh{x'}) e^{\frac{\beta \hslash^2 \mu^2}{2m}}
\end{eqnarray}
Now to find the kinetic energy, we first do the double differentiation as done earlier, and then substitute $x=x'$, integrate with respect to $x$ as we have done earlier. Therefore, we get from Eq. \ref{ke_int}.
\begin{equation}
    I_{2} =  \sum_{\mu=1}^{\lambda} (N_{\lambda}^{\mu})^2\left( \frac{\mu(\lambda + \mu)!}{(\lambda - \mu)!}- \frac{2\lambda (\lambda +1)}{2\lambda+1} \frac{(\lambda+\mu)!}{(\lambda-\mu)!} \right) e^{\frac{\beta \hslash^2 \mu^2}{2m}}
\end{equation}
Thus, the complete expression for  the interaction term is: 
\begin{equation}
    = (e^{-\beta W_{0}}-1) \sum_{\mu=1}^{\lambda} (N_{\lambda}^{\mu})^2\left( \frac{\mu(\lambda + \mu)!}{(\lambda - \mu)!}- \frac{2\lambda (\lambda +1)}{2\lambda+1} \frac{(\lambda+\mu)!}{(\lambda-\mu)!} \right) e^{\frac{\beta \hslash^2 \mu^2}{2m}}
\end{equation} 
The kinetic energy after performing the inverse Laplace transform, the third term will also give the same resultant; therefore, the final expression is 
\begin{equation}
    T_{int}[\rho] = -\frac{1}{2}\frac{\hslash^2}{2m} \sum_{\mu=1}^{\lambda} \left( \mu^2- \frac{2\lambda  (\lambda +1)\mu}{2\lambda+1} \right) \left( \Theta(\ef -W_{0}+\frac{\hslash^2 \mu^2}{2m} ) -\Theta(\ef+ \frac{\hslash^2 \mu^2}{2m}) \right)
\end{equation}

\section{The actual kinetic energy of P\"oschl-Teller potential}
%\noindent For the ground state, the solution reduces to the regular associated Legendre polynomial (no nodes) \(\mu=\lambda-n\):
%\[
%(\psi_{\lambda}^{\mu})_0(y) = N_{\lambda}^{\mu} P^{\mu}_{\lambda}(\tanh{x}),
%\]
%where \(N_{\lambda}^{\mu}\) is the normalization constant. 
%Normalization requires:
\[
\int_{-1}^1 |(\psi_{\lambda}^{\mu})_0(y)|^2 \frac{dz}{ (1 - y^2)} = 1 \implies  (N_{\lambda}^{\mu})_0^2= \frac{\mu (\lambda -\mu)!}{(\lambda +\mu)!} = \frac{1}{2}.
\]
Where $(N_{\lambda}^{\mu})_0$ is the ground state normalization constant for the P\"oschl-Teller potential.
\subsection{The Expectation value of Potential Energy}
\[
\langle V \rangle = \int_{-1}^1 (\psi_{\lambda}^{\mu})_0(y)^2(y) V(y) \frac{dy}{ (1 - y^2)}.
\]
Substituting \(V(y) = -\frac{\hslash^2 }{2m} \lambda(\lambda + 1)(1 - y^2)\) and \(\psi_0(y)\):
\[ \langle V \rangle = -\frac{\hslash^2 \lambda (\lambda+1)}{2m} (N_{\lambda}^{\mu})^2\int_{-1}^{1} (P_{\lambda}^{\mu}(y))^2 dy \]
Using the Identity \( \int_{-1}^{1} P_{\lambda}^{\mu}(x) P_{\lambda'}^{\mu}(x) \, dx = \frac{2}{2\lambda+1} \frac{(\lambda+\mu)!}{(\lambda-\mu)!} \delta_{\lambda \lambda'}
\)\\
\begin{equation}
    \langle V \rangle = -\frac{\hslash^2}{2 m}  \frac{2\lambda^2(\lambda+1)}{(2\lambda+1)}
\end{equation}
\subsection{Expectation value of Kinetic energy}
The total ground state energy is
\( E_{0} = -\frac{\hslash^2 \lambda^2}{2m}\)\\
Using \( \langle K \rangle = \langle E_{0} \rangle - \langle V \rangle\):
\begin{equation}
    \langle K \rangle = \frac{\hslash^2}{2m}\frac{\lambda^2}{(2\lambda+1)} 
\end{equation}

\section{The interaction potential term for different values of $M$ in Suzuki-Trotter decomposition}
\begin{equation}
h_{int}(x'', x') =    \bra{x''}\left( \lim_{M\longrightarrow \infty} \left(e^{-\frac{\beta}{M} \ham_0 }e^{\frac{\beta}{M}\hat W} \right)^M \right) \ket{x'} - \bra{x''} e^{-\beta \ham_{0}}\ket{x'}
\end{equation}
\begin{enumerate}
    \item $\mathbf{M=1}$
\begin{equation}
     h_{int}(x'', x') = (- 2 \pi \beta g^2 x')G^{0}(x'', x'; \beta)
 \end{equation}
 \item $\mathbf{M=2}$
 \begin{equation}
     h_{int}(x'', x') = ( -2 \pi \beta g^2 x'+ \frac{\pi^3 \beta^3 g^4 \hslash^2}{2m}) G^{0}(x'', x'; \beta)
 \end{equation}
 \item $\mathbf{M=3}$
\begin{align}
    h_{int}(x'',x')= \bigg[-2\pi x'\beta g^2+ \frac{64}{3^4}\pi^2\beta^3g^4\frac{\hslash
    ^2}{2m}+\frac{96}{3^5}\pi^3\beta^4g^6\frac{\hslash^2}{2m}x'\bigg]G^0(x'',x';\beta)
\end{align}\textbf{}
\end{enumerate}
\end{appendix}
\bibliography{TCA}
\end{document}